\documentstyle[preprint,aps]{revtex}

\begin{document}
\draft
\title{
Temperature dependent gap anisotropy in Bi$_{2}$Sr$_{2}$CaCu$_{2}$
O$_{8+x}$ as evidence for a mixed-symmetry ground state}
\author{J. Betouras and R. Joynt}
\address{
Department of Physics and Applied Superconductivity Center\\
University of Wisconsin-Madison\\
1150 University Avenue\\
Madison, Wisconsin 53706\\}
\date{\today}
\maketitle
\begin{abstract}
In a recent experiment, Ma et al. measured the temperature dependence
of the gap anisotropy of oxygen-annealed Bi$_{2}$Sr$_{2}$CaCu$_{2}$O$_{8+x}$.
Their measurements were taken along the two directions $\Gamma - M$ and
$\Gamma - X$. They found that the gap along both directions
is nonzero at low temperatures and that the ratio is strongly temperature
dependent.  We show, using Ginzburg-Landau theory,
that this behavior can be obtained if one assumes the existence of s-wave
and d-wave components for the order parameter.  Our theory predicts
orthorhombic anisotropy in the gap and
anomalous behavior for the electronic specific heat
below T$_c$.
\end{abstract}
\narrowtext

The most central issue in high-T$_c$ superconductivity
at the present time is the nature of the order parameter.
Many experiments have attempted to clarify the form
and symmetry of the order parameter, with the most
conclusive being of the phase-sensitive tunneling variety
\cite{woll,chaud,brawn}. These indicate that the gap function
changes sign when rotating by $\pi/2$ in momentum space.
This suggests a d-wave order parameter in the system studied,
namely YBa$_2$Cu$_3$O$_{7-\delta}$ with small $\delta$.
Spin-fluctuation models of high-$T_c$ lead to d-wave
superconductivity \cite{pines,scala}. Strong correlation models have a similar
instability to d-wave pairing
\cite{gros,lee}. Thus the experiments on
YBa$_2$Cu$_3$O$_{7-\delta}$ may be taken as a confirmation
of the magnetic mechanism in general. To distinguish between the two
models is not an easy problem. However, variational Monte
Carlo (VMC) calculations on a strong correlation model, the extended
t-J model, have suggested that, while a d-wave is the dominant instability,
s-wave and d-wave mixing may be the favored ground state for higher
doping levels \cite{li} and certain band structure parameters
\cite{kolt}. From this point of view, not all high-$T_c$
systems are alike, and it becomes crucial to perform experiments
on different systems and, most importantly, at different doping levels.

Apart from the tunneling technique, the most direct measurement (although
not phase-sensitive) is angle-resolved photoemission (ARPES).
A recent experiment by Ma et al. \cite{ma} used ARPES to investigate the
gap in Bi$_{2}$Sr$_{2}$CaCu$_{2}$O$_{8+x}$
single crystals. In particular this
group has measured the gap at two points on the Fermi surface:
$\Delta$($\Gamma - M$) or $\Delta$($k_{x}=0$)
in our coordinate system (along the Cu-O bond)
and $\Delta(\Gamma -X)$ or $\Delta(k_{x}=k_{y})$.
This was done as a function of temperature.
The basic features of the experimental results are :
(i) the gap $\Delta(\Gamma - M)$ opens up quickly below T$_c$ and
$\Delta(\Gamma - M) > \Delta(\Gamma -X)$
for all temperatures;
(ii) $\Delta(\Gamma - X)$ is small or even zero for
T $>$ 0.8 T$_c$ and then increases gradually as T is further reduced,
leading to a strongly temperature-dependent gap anisotropy;
(iii) there are indications of upward curvature in
$\Delta$($\Gamma -X)$;
(iv) the temperature dependence of $\Delta$ is weak in both
directions for a wide range of
temperatures (from T=0 K to T=0.7 $T_{c}$ approximately).
These data are different from results obtained by other workers
on Bi$_{2}$Sr$_{2}$CaCu$_{2}$O$_{8+x}$ \cite{shen,ding} in that the
gap along $\Gamma-X$ is not always small. This is due to annealing in an oxygen
atmosphere, leading to higher oxygen
content and the presumably to a higher hole doping level in the CuO$_2$ planes.
\cite{ma}

We do not possess a detailed microscopic theory of s-d mixing
and the VMC calculations are restricted to T=0.  Thus
we work within the framework of Ginzburg-Landau theory.
The point group symmetry of the CuO$_2$ planes in the temperature
range of interest is orthorhombic $C_{2v}$, due to buckling of
the planes \cite{kirk}.
In this symmetry the interesting thing is that an order parameter with
d-wave symmetry transforms according to the identity representation.
[These symmetry issues are discussed in detail in Ref.\cite{li}].
So from a strict symmetry point of view a d-wave and an
s-wave are indistinguishable.
The concept of a two-component order parameter, where  one component
has an s-wave symmetry and the other a d-wave symmetry is thus
well-defined
only in the tetragonal case.  In the present orthorhombic
case we can regard the s-d mixing as a strong temperature
dependent anisotropy which nevertheless never actually
breaks the crystalline symmetry. If we write the
Ginzburg-Landau free energy density as follows:

\begin{eqnarray}
f=&&\alpha_s^0(T-T_s){|\Psi_s|}^2 +\alpha_d^0(T-T_d){|\Psi_d|}^2+
\beta_s{|\Psi_{s}|}^4 +\beta_d{|\Psi_{d}|}^4+
\beta{|\Psi_{s}|}^2{|\Psi_{d}|}^2 \nonumber\\
&&+\beta_{sd}\left({\Psi_{s}}^2
{\Psi_d}^{*2}+ c.c. \right) +\gamma\left(\Psi_s{\Psi_d}^* + c.c.\right),
\end{eqnarray}
the difference from the case of tetragonal symmetry is the addition of the
final
bilinear term. In fact the coupling constant $\gamma$ is proportional
to the orthorhombic distortion $b-a$ where b and a are the
lattice constants in the Cu-O plane.
Note that in our case, the gradient terms are not necessary
due to the absence of external fields and the assumed
homogeneity of the material
(translational invariance).
We minimize the free energy  with
respect to $|\Psi_{s}|$ and $|\Psi_{d}|$ and the phase difference
of the two components $\phi$. The phase difference is crucially
important. If it is zero then the resulting state is s+d: the gap function
is real (except for the overall phase related to gauge freedom). If
$\phi$ is nonzero the state is more similar to  s+id: it breaks time
reversal symmetry and has no nodes.

The minimization leads to coupled equations which we have solved
numerically.
The result of the calculations is that, according to the
sign of the parameter $\beta_{sd}$ in the
above functional, there are two possibilities
for the value of $\phi$. It is either zero  or there is a transition
from a continuous value (temperature dependent) to zero.
We attempted to fit the data with a temperature dependent phase shift
without success. The data therefore point to the s+d state.
So we adopted the case $\phi=0$
for the phase difference between the two components of the
order parameter.  We chose  $\beta_{sd}<0$ (which
then essentially adds to $\beta$) and $\gamma<0$.
The equations we have to solve in
order to get the order parameters become:

\begin{eqnarray}
\alpha_{s,d} \vert \Psi_{s,d} \vert +2\beta_{s,d} \vert \Psi_{s,d} \vert ^3 +
\left(\beta+\beta_{sd}\right)|\Psi_{d,s}|^2|\Psi_{s,d}| +
{\gamma \over 2} \vert  \Psi_{d,s} \vert =0 .
\end{eqnarray}

The gap function  we consider is of the form:
\begin{eqnarray}
\Delta({\bf k})=|\Psi_d|\: f_d({\bf k}) + |\Psi_s|\: f_s({\bf k}) ,
\end{eqnarray}
where {\bf k} lies on the Fermi surface.
The assumed forms of the k-dependence are :
\begin{eqnarray}
f_d(k_x=k_y)=0  ,  f_d(k_x=0,k_y)=1\\
f_s(k_x=k_y)= f_s(k_x=0,k_y)= 1
\end{eqnarray}

Accordingly, the  gap along the $\Gamma- X$ direction is
the solution $|\Psi_{s}|$ whereas along the $\Gamma - M$ direction it is
$|\Psi_{s}|+|\Psi_{d}|$.
In order to compare the data over the whole temperature range
to experiment we used
the approximation $\alpha_{s,d} / \beta_{s,d} \propto [\tanh (\nu ({T_{c}
\over T}-1)
^{1/2})]^2$.  For $\nu = 1.74 $ this form gives a good fit to
the weak-coupling BCS temperature dependence of the gap.
 Increasing $\nu$
is a way of mocking up strong-coupling effects.
In Fig.\ \ref{eva}  we show the results of the fitting to experimental data.
Several features of the fit are noteworthy.
A strong-coupling value of the parameter $\nu$
must be used in order to represent feature (i) mentioned above.
T$_s$ must be lower than T$_d$ by about 15 K so that the s-component
is suppressed near T$_c$, feature (ii) above.
The upward curvature in  $\Delta(\Gamma -X)$ [feature (iii)] is only
produced by having a nonzero value of $\gamma$, the bilinear
coupling of the two components.
Finally, the tangent function in the coefficients of the
quadratic terms in the free energy produces a temperature
independent gaps at low T, as in ordinary BCS theory,
[feature (iv)]. There is only one second order transition.

Near $T_{c}$ the gap is pure d-wave, with gap zeros in the diagonals
and maxima along the axes. As the temperature is reduced, the s-wave
component grows, the nodes  move towards the $k_{y}$ axis and the
gap along this axis become smaller while the gap along the $k_{x}$
axis becomes larger. Thus it is a temperature-dependent orthorhombic
anisotropy. We take, for illustration purposes only, a circular Fermi surface
with
$f_d=\cos(2\theta_{k})$ , $f_s=1$. Then the
zeros of the gap move from
a value of $ \theta_{k}= \pi / 4 $ close to $T_c$ to a value
which is temperature
dependent ( $ \theta_{k}=-1/2 \arccos (-\Psi_s/ \Psi_d) $ ).
This is shown in Fig.\ \ref{duo}. In this extreme example the zeros of the gap
move all the way to the $k_{y}$ axis and then disappear. We stress
that this disappearance is a highly model-dependent phenomenon,
depending, for example on the detailed momentum dependence of the
s-wave component. On the other hand, the qualitative result that
the zeros of the gap will
move as the temperature is varied is a firm prediction of the two-component
theory as long as the components have different temperature dependences,
as is indicated by the experiments. Note also that experiments such as
the reported in Ref.1 which depend only on the relative phase change
under a $\pi/2$ rotation in k-space and  will not be affected by a small
movement of the nodes.

In Fig.\ \ref{tria} , the anisotropy ratio
$\Delta(\Gamma-X)/\Delta(\Gamma-M)$
is plotted as a function of temperature.
The dramatic temperature variation of this quantity is well
reproduced by the strong-coupling parameters. Also in Fig.\ \ref{tessera},
the form of the heat capacity (in arbitrary units) is plotted.
Since we are interested for values close to $T_c$ , we used  the
linear form of the parameters $\alpha_s$ and $\alpha_d$ , in
such a way that the  first term of the Taylor  expansion
(with $T_c$ instead of T
in the denominator)  of the hyperbolic
tangent agrees. The hyperbolic tangent form was chosen above
 in order to compare the data  with the theory
at much lower temperatures. A somewhat anomalous behaviour
of $C_{v}(T)$ is predicted, with a local
maximum around  0.7 $T_c$. This comes from the sharp growth  of the second
component in the order parameter in this temperature range.

We do not claim that the fit in Fig.\ \ref{eva} is a quantitative
confirmation of the two-component theory. The curves are smooth
and the numbers of
parameters large. It is true, however, that the temperature-dependent
anisotropy of the superconducting gap in BSCCO
is what one expects if  a two-component order parameter
exists. A pure  d-wave order parameter alone
cannot account for the data. Our main point in this
paper is to note that the two-component picture has a number of
surprising consequences which may be checked in highly hole-doped
BSCCO. These are : the gap energy should have orthorhombic anisotropy
at low temperatures; there should be upward curvature in $\Delta(\Gamma-X)(T)$
already suggested by the existing data; the position of the gap nodes
should be temperature-dependent; there should be a peak in the
electronic specific heat close to 0.7 $T_c$.

We would like to thank Marshall Onellion, Ron Kelley, Jian Ma,
Christoph Quitmann
for valuable discussions. This work was supported by the National Science
Foundation under grants No. 9214739 and No. 921047.

\eject

\begin{figure}
\caption{Shows the gap along the two directions of the Fermi surface in
comparison to the experimental data. The solid line is the solution
for $\nu=3.0$ (the values of the parameters are :
$\alpha_{s}/\beta_{s}=
-250 [\tanh(3.0 (0.8 T_{c}/T-1)^{1/2})]^{2}$,
$\gamma/\beta_{s}=\gamma/\beta_{d}=-100$,
$\beta/\beta_{s}=\beta/\beta_{d}=2$,
$\alpha_{d}/\beta_{d}=-220 [\tanh(3.0 (T_{c}/T - 1)^{1/2})]^{2}$,
the dashed line is the solution for $\nu=1.74$
(the values of the parameters are:
$\alpha_{s}/\beta_{s}= -340 [\tanh(1.74 (0.8 T_{c}/T - 1)^{1/2})]^{2}$,
$\gamma/\beta_{s}=\gamma/\beta_{d}=-100$, $\beta/\beta_{s}=\beta/\beta_{d}=
2$, $\alpha_{d}/\beta_{d}=-280 [\tanh(1.74 (T_{c}/T - 1)^{1/2})]^{2}$).}
\label{eva}
\end{figure}

\begin{figure}
\caption{Shows the zeros of the gap as a function of temperature, for
the hypothetical circular Fermi surface. At
low temperatures there are no zeros in the gap according to the  model
and $\Psi_s$ is greater than $\Psi_d$.}
\label{duo}
\end{figure}

\begin{figure}
\caption{Shows the anisotropy ratio defined as $\Delta(\Gamma-X)/
\Delta(\Gamma-M)$ for the value of $\nu=3.00$.}
\label{tria}
\end{figure}

\begin{figure}
\caption{The specific heat as predicted by the model. Near 0.7 $T_c$
there is a local maximum.}
\label{tessera}
\end{figure}
\end{document}